\documentclass[sigconf]{acmart}

\copyrightyear{2022} 
\acmYear{2022} 
\setcopyright{acmcopyright}\acmConference[NordiCHI '22]{Nordic Human-Computer Interaction Conference}{October 8--12, 2022}{Aarhus, Denmark}
\acmBooktitle{Nordic Human-Computer Interaction Conference (NordiCHI '22), October 8--12, 2022, Aarhus, Denmark}
\acmPrice{15.00}
\acmDOI{10.1145/3546155.3546669}
\acmISBN{978-1-4503-9699-8/22/10}

\begin{document}

\title{Assessing Peer Award Diversification on Reddit}

\author{Amaury Trujillo}
\orcid{0000-0001-6227-0944}
\affiliation{%
	\institution{IIT-CNR}
	\city{Pisa}
	\country{Italy}}
\email{amaury.trujillo@iit.cnr.it}




\begin{abstract}

Monetizing user-generated content in social media such as Reddit, in which users are both content creators and consumers, is challenging.
Among the platform's strategies we find Reddit Awards, paid tokens of appreciation given by peer users who markedly enjoyed a particular posted content. Initially, there were only three awards, but the platform later greatly expanded their number and variety.
This work thus aims to investigate how awarding changed after such diversification.
To this end, two datasets of posts made before and after the change (16M submissions and 203M comments) were analyzed by operationalizing awarding level and award diversity.
Results show that after diversification, the awarding level increased across multiple measures, both significantly and considerably, albeit two of the original three awards remained by far the most commonly given.
Such an increase indicates that providing more award options benefits users and is a viable user interaction approach for platforms to both engage users and monetize their content.

\end{abstract}

\newcommand{\dsname}[1]{\texttt{\small{#1}}} % dsname -- type dataset names for 2019H1 and 2020H1
\newcommand{\award}[1]{\textit{#1}} % award name 
\newcommand{\subr}[1]{\texttt{\small{r/#1}}} % subreddit name
\newcommand{\redditor}[1]{\texttt{\small{u/#1}}} % redditor name
\newcommand{\ptp}{{\small{PTP}}} % per thousand posts acronym
\newcommand{\mad}[1]{({\small{MAD}}=#1)} % median absolute value

\maketitle

\section{Introduction}

Maintaining an engaged user base and monetize it has always been paramount but very demanding for social media platforms~\cite[chap. 1]{faltesek2018selling}.
In this regard, some platforms have established content creator schemes in which approved users can monetize their content based on the engagement of content consumers.
However, such schemes present an imbalanced power dynamic between platform and creators~\cite{kopf2020rewarding}, and in platforms in which there is no clear distinction between content creators and consumers, e.g., online discussion forums, these are not a viable option.
Reddit, one of the most visited websites worldwide,\footnote{\url{https://www.alexa.com/siteinfo/reddit.com}} is a representative example.
After building an initial critical user base, in 2009 the platform turned to the commonly used and abused advertisement business model. However, as the number of users continued to rapidly grow, thanks to engaging content and discussions,  Reddit's operation became hard to sustain.
Hence, in July of 2010, the platform introduced Reddit Gold, a paid subscription membership with exclusive features, as an additional revenue stream to cover its ever increasing operating costs. In November of 2012, Reddit Gold was expanded with the ability to give other users gold features for a limited time, via the \emph{gilding} of posts; it was the platform's precursor of a monetized peer award system.

\begin{figure}
   \centering
   \includegraphics[width=1\columnwidth]{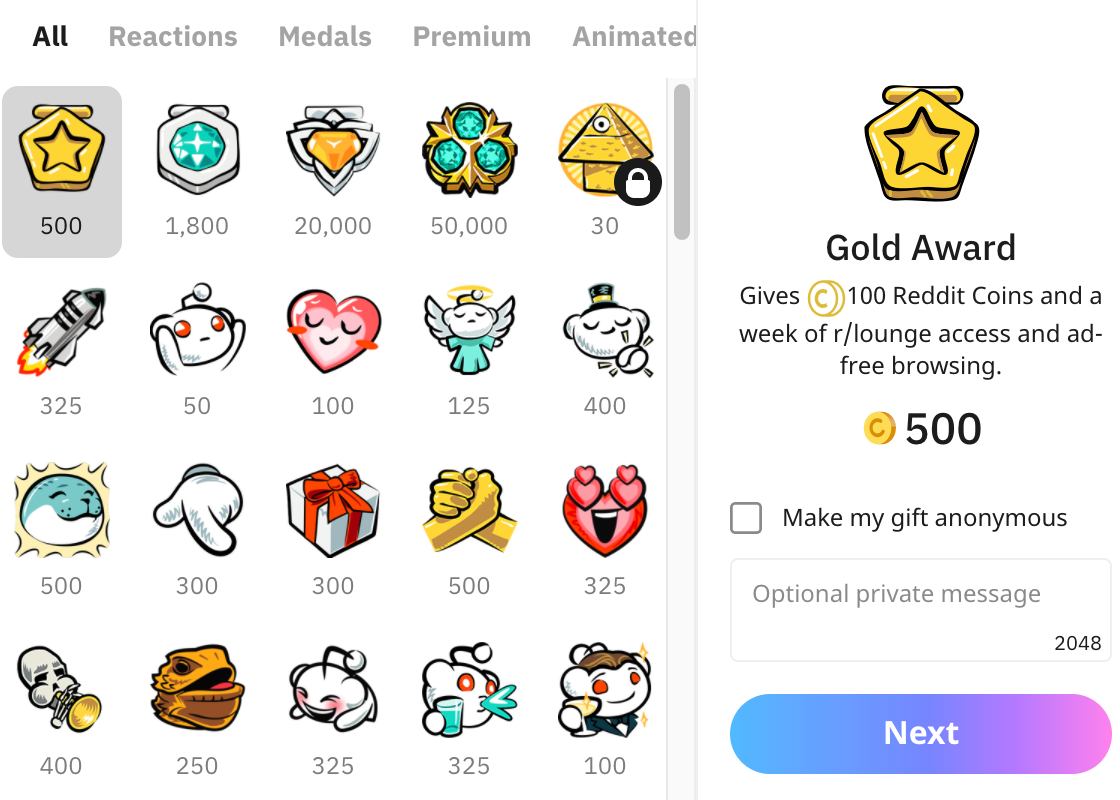}
   \caption{Awarding modal dialog on Reddit's web app.}
   \Description{}
   \label{fig:award_modal}
\end{figure}

Six years later, in September of 2018, the platform introduced Reddit Coins, a new monetization strategy, with the platform's paid subscription membership being renamed from Reddit Gold to Reddit Premium.
Reddit Coins\footnote{\url{https://www.reddit.com/coins}} are virtual goods that can be purchased by users ---with or without Reddit Premium--- to be exchanged by \emph{awards} that are used as tokens of appreciation for submissions or comments posted by other users.
Some Reddit awards grant the recipient benefits, such as coins or access to exclusive features within the platform; other awards function merely as a decoration for the recipient's post.
Initially, only three awards were available: the original \award{Gold}, plus \award{Silver} and \award{Platinum}; although giving any of these was still called \emph{gilding}.
Then, in July of 2019, Reddit introduced Community Awards (exclusive to a given subreddit) and significantly expanded the general awards from three to dozens (and counting), thus diversifying awards platform-wise.
The new awards, often depicted with colorful imagery (see Figure~\ref{fig:award_modal}), are mostly humorous references to Internet slang, memes, or inside jokes. Consequently, the act of giving an award was renamed from gilding to \emph{awarding}, albeit the core mechanism remained the same.

Obviously, the main goal of the diversification was to increase the level of awarding and thus increase revenue from the direct selling of Reddit Coins and/or the Reddit Premium subscription that includes a fixed monthly amount of Coins plus access to exclusive awards. However, despite the benefits in user interaction touted by Reddit personnel and the enthusiasm of many users, others decried these changes, mainly for its focus on monetization and perceived lack of appeal when compared to the then-current gilding mechanism,\footnote{\url{https://www.redd.it/chdx1h/}} as expressed by redditor \redditor{Poiuy2010\_2011}:

\begin{quote}
Maybe I'm in the minority here but I find these community awards absolutely useless. With the standard 3 awards there is a clear hierarchy. But with the community awards, especially when a post gets popular, all of them just kinda blend together and ultimately become meaningless.
\end{quote}

%TODO: add more information regarding the research questions to address AC review suggestions

In this context, the present work aims to investigate the impact in awarding behavior of expanding awards beyond gilding within Reddit, around the following research questions (RQ):

\begin{itemize}
	\item RQ1: How award diversification changed awarding levels?
	\item RQ2: How diverse are the awards given by Reddit users?
\end{itemize}

Results indicate that indeed awarding levels (based on different metrics) increased after diversification; still, the original and more hierarchical gilding awards remained the most popular awards, particularly \award{Silver} and \award{Gold}. Hence, this study offers two primary contributions: an operationalization of appreciation token levels and diversity, and suggestions for the implementation of similar paid appreciation token schemes in other platforms based on peer user-generated content.

\section{Background and Related Work} \label{sec:background}

Many social media platforms have developed virtual currencies as a potential source of revenue stream, although with varying degrees of success, as exemplified by two of the biggest platforms.
On one hand, back in 2005 Tencent introduced to great success QQ Coins as a means to pay for online services and virtual goods within its ecosystem~\cite{gans2015some, halaburda2016beyond}, which with the years expanded to various in-platform financial activities~\cite{zhou2017proguard}, and is still widely used by hundreds of millions of users.
On the other hand, in 2011 Facebook Credits were made available to purchase digital items within the Facebook platform, subject to transaction fees, but the currency was phased out two years later after user and developer complains~\cite[chap.~3]{halaburda2016beyond}, as well as more rigid financial regulation~\cite{gans2015some}.
Incidentally, Facebook also intended to develop its own cryptocurrency, called Diem (formerly known as Libra), but it was promptly met with government scrutiny and public mistrust~\cite{chiu2021should}, with the company getting rid of the project in January of 2022, before any launch.

Arguably, the aforementioned increasing legal probes on virtual currencies compelled Reddit to explicitly use the term \emph{virtual goods} to refer to its own monetization scheme, despite that the name Reddit Coins is more evocative of currency.
The same term is also used in a similar monetization scheme that the streaming platform Twitch introduced in mid-2016, called Cheering with Bits.
To show support to streamers, Twitch users can purchase virtual goods called Bits, with which they Cheer on a given stream channel chat, earning a channel-specific badge representing the amount of Bits donated. Twitch offers a few common badges related to Bits, but affiliates and partners can replace the badges in specific channels, thus diversifying the imagery in a manner reminiscent of Reddit Community Awards. TikTok has a similar yet different monetization mechanism for its streaming platform, called Gifting. Users buy TikTok Coins ---which the platform explicitly treats as \emph{virtual items} and not property in its terms of service--- in order to purchase in-platform virtual Gifts to donate to live streamers. The Gifts are then transformed into Diamonds, which are valued proportionally to the gift coin value (minus the platform's share) and can be then redeemed for money by streamers.

Twitch Bits have challenged the dominance of third-party donation tools~\cite{partin2020bit}, and ---together with paid subscriptions--- they have become one of the main mechanisms with which Twitch streamers monetize their content~\cite{johnson2019and}, and users manifest their appreciation and financial commitment to streamers~\cite{wolff2022audience}. TikTok Coins are also considered to be key in the financial success of TikTok and some of its content creators~\cite{mhalla2020video}.
Hence, Twitch Bits and TikTok Coins share more of the characteristics of currency as a medium of exchange than Reddit Coins.
In fact ---and in spite of their designation as virtual \emph{goods} or \emph{items} by their creators--- in the literature all three are treated as in-platform currency~\cite{johnson2019and,partin2020bit,hartzell2021source,wolff2022audience, domingues2021analyzing}. However,  they do not act as a store of value and are limited as a medium of exchange or unit of account, which are generally considered to be the functions of currency~\cite{halaburda2016beyond}.
Hence, I refer to these and other similar monetization virtual objects as \emph{paid appreciation tokens}: recognition symbols of the enjoyment that some content has brought to their giver, who had the willingness to pay for them.

There are, however, three fundamental differences between these video streaming platforms and Reddit that affect their paid appreciation token mechanisms.
First, the content dynamics are different: original content creation is a core aspect of Twitch and TikTok, while content sharing and discussion is central in Reddit, irrespective of whether such content was originally created by the poster or not.
Second, streamers monetize their content whenever they receive Cheers or Gifts, earning real-world money proportional to the donated Bits or Coins, while Reddit users do not receive financial compensation when their post is awarded; at most, some awards include access to exclusive Premium features or Reddit Coins to be further exchanged by other awards. After all, it is the appreciation token that has been paid, not necessarily the creator or sharer of the content to which the token was bestowed.
Third, the relationship between the parties involved is quite different. In Twitch, interactions revolve around channels, managed by content creators (who often hope to earn money for their effort) for content consumers (who sometimes monetarily reward such efforts). TikTok is similar, but there are no channels, the interaction is directly with content creators. In Reddit, the user-generated content process revolves around topical communities, in which almost any registered Reddit user can act as both creator and consumer, without the expectation of earning money for posting content. In other words, users are peers, hence the term \emph{peer awards} is also used to designate Reddit's paid appreciation tokens.

Notwithstanding these distinctive features of Reddit and its popularity, there is scarce literature on its peer awarding mechanisms, old and new. The majority of academic work on Reddit that takes into consideration the awarded (or gilded) status does so as a content feature for a specific task, such as the modeling of content popularity~\cite{wigsnes2019predicting} or user personality~\cite{gjurkovic2018reddit}. 
The earlier of a couple works focused on Reddit awarding concerns an analysis of pre-Coins gilded content from a sample of 25M comments from the top 100 subreddits by content in May of 2015~\cite{mendelsohn2017giving}.
However, due to the focus on linguistic features, the work only considered comments, excluding submissions (which could represent links or embedded multimedia).
Therein, authors found that subreddits could be clustered in different groups based on linguistic and other comment features, as well as the topic of interest.
In most cases, the initial comments (relative to their submission) are more likely to be gilded compared to late comments, lengthier comments are preferred to shorter ones, and that the subreddit clusters had different preferences of writing style, e.g., the use of ``you'' and a more narrative style for sports-related communities.

A more recent and in-depth analysis concerns the post-Coins gilding process. In this experimental study by \citeauthor{burtch2021peer}~\cite{burtch2021peer}, \emph{Gold} was randomly and anonymously given to 905 users' posts over a two-month period within three writing-focused subreddits (\subr{hfy}, \subr{nosleep}, and \subr{shortscarystories}). Therein, authors found that peer awards induced recipients to produce longer and more frequent content, especially among newer community members, with such posts being textually similar to their own past awarded content. Hence, besides the monetization itself of user-generated content, we could say that peer awards also motivate users to create more content, which in turn attracts new users to the platform or retains the attention of those already present ---producing a positive network effect--- while also increasing the chances of new content being awarded. That being said, the experimental study in question was limited to a single award kind, with a very small number of subreddits, and a relatively small quantity content, which is expected to have a very specific narrative nature. Further studies are necessary to better understand peer paid appreciation tokens.

In that regard, and as far as I am aware, the present work is the first academic work on the new Reddit's awarding mechanism in particular, and the first quantitative study on the diversification of paid appreciation tokens on social media platforms in general.

\section{Data} \label{sec:data}

Subreddits represent communities of users who share a common interest, each developing its own specific identity, participants, and dynamics.
Hence, in order to answer the research questions, both at the platform and community levels, the first step was to select a representative sample of subreddits.
To this end, I selected the top 50 subreddits by number of subscribers as of January 1, 2019, with the following exclusion criteria: 1) it is an official subreddit of Reddit (not user-driven); and 2) it is marked as adult content, i.e., \emph{not safe for work} (NSFW).
Only two subreddits were excluded based on these criteria, which were respectively: \subr{blog}, the official Reddit's blog subreddit with 16.8M subscribers, most by default upon registration; and \subr{gonewild}, a subreddit for the exchange of nude and sexually explicit photos with 1.7M subscribers.

Concerning Reddit posts, there are two kinds: submissions and comments.
A submission is posted directly to a given subreddit, it has a title, and either a text body or a link (e.g., to a website or an embedded multimedia file).
A comment represents textual content and is posted in response to a given submission or to another comment. Both submissions and comments can receive awards, but their dynamics are intrinsically different~\cite{weninger2014exploration}.
Hence, awarding is analyzed separately for each kind.
For the retrieval of submissions and comments for the selected subreddits, I used the monthly dumps of the Reddit Dataset from Pushshift, a platform for the collection, analysis, and archiving of several social media \cite{baumgartner2020pushshift}.

Regarding the data timeframe, less than one year passed between the introduction of the gilding and awarding mechanisms, with both being introduced on the third quarter, albeit neither mechanism was widely adopted immediately.
Therefore, I collected the subreddits' submissions for a period of six months for each mechanism starting from the calendar year following its introduction ---thus excluding the first months of adoption--- which corresponds respectively to the first halves of 2019 and 2020.
Henceforth these two periods and their respective datasets will be referred to as \dsname{2019H1} for the gilding and \dsname{2020H1} for the diversified awarding.
Concerning the comments, those made within the following two months (60 days) after their submission's creation were retrieved for every submission in the sampling.
The vast majority of Reddit comments are made within the first few hours of a submission~\cite{medvedev2017anatomy}, thus this timespan should cover all of the comments made to a submission, save for a few exceptions.
In \dsname{2019H1}, there are 7.47M submissions (6.9\% of the grand total) and 100.48M comments. In \dsname{2020H1}, there are 8.38M submissions (5.4\% of the grand total) and 103.04M comments.
Finally, the awards received, if any, for each submission and comment were extracted.
It should be noted, however, that the identity of the giver and the time of awarding are not available.
Hence, in the case of \dsname{2019H1}, for the analyses only the three gilding awards are taken into consideration, with any new non-gilding awards (most likely given during the transition period) being ignored.

\section{Methods}
 
First, I operationalized the research questions by defining objective measures of awarding level (RQ1) and award diversity (RQ2); then, I chose adequate statistical methods for these based on a preliminary analysis; finally, I conducted the respective data analyses by dataset, subreddit, and post kind, as detailed in the following paragraphs.
 
\subsection{Measurement of Awarding Level}
 
To describe the level of awarding and its possible increase after award diversification, the number of awards given seems to be the most straightforward measure. However, this measure would offer an insufficient view, as it is possible that the awards given increase but that the number of coins spent for them does not, i.e., more but cheaper awards are given. In addition, perhaps there is an increase in both awards and coins, but the proportion of awarded posts remains stable, i.e., the same proportion awarded of posts receives more awards and higher coin-value. Hence, for a more comprehensive description of awarding level (RQ1), I use the following three measures:

\begin{itemize}
	\item \emph{Awards given}: the count of awards given, regardless of their identity; also referred to as awardings.
	\item \emph{Coins spent}: the sum of distinct award prices multiplied by the respective number of times it was given.
 	\item \emph{Awarded posts}: the number of posts that received at least one awarding.
\end{itemize}

In order to have standardized measures while comparing results ---and given the relative small shares of awarded content--- henceforth all of the above measures are expressed in \textit{per thousand posts} (\ptp). At the subreddit level, dispersion is measured with the median absolute deviation (MAD), and to test the paired differences of awarding measures between \dsname{2019H1} and \dsname{2020H1}, I used two-sided Wilcoxon signed-rank tests. Effect sizes are also given, both as percentage growth and paired median differences. For the latter, a 95\% confidence interval (CI) is calculated via DABEST (data analysis with bootstrap-coupled estimation)~\cite{ho2019moving}, based on 5000 resamples.

\subsection{Definition of Award Diversity}

The definition and measurement of diversity have been the source of much terminological confusion  ~\cite{hill1973diversity, jost2006entropy, tuomisto2010consistent}.
This is particularly the case in Ecology, in which the study of species diversity has been traditionally described with several diversity indices, such as Richness, Shannon index, and Gini-Simpson index~\cite{tuomisto2010consistent, roswell2021conceptual}. 
However, these indices measure different things: Richness is a count of distinct types (e.g., species), the Shannon index is an entropy, and the Gini-Simpson index is a probability~\cite{tuomisto2010consistent}. In a seminal work~\cite{hill1973diversity}, Hill presented a formal discussion on the
concept of species diversity and described it as the inverse of mean species proportional
abundance; further, Hill also defined a unifying notation in which different means ---harmonic, geometric and arithmetic--- correspond to the aforementioned traditional diversity indices. However, the Hill diversity only became widely known decades later, after the publication of a highly influential scientific opinion piece by \citeauthor{jost2006entropy}~\cite{jost2006entropy}, in which it was called the ``true diversity''.
In its most common definition~\cite{tuomisto2010consistent} ---and substituting species by awards--- the Hill diversity can be written as:

\begin{equation}
\label{eqn:hill_diversity}
^qD = \left( \sum_{i=1}^A p_i^q \right)^{1/1-q}
\end{equation}

In other words, the Hill diversity measures the mean rarity of awards in a sample, while also indicating that a community with rarer awards (on average) has higher diversity~\cite{roswell2021conceptual}.

Based on a preliminary analysis of \dsname{2020H1} there are few abundant awards (namely \emph{Silver} and \emph{Gold}) and many rare awards, thus setting $q$ to 1 is a sensible choice. In this case, we can use the more known Shannon index,

\begin{equation}
	H = -\sum_{i = 1}^A p_i \ln(p_i)
\end{equation}

and rewrite Eq. \ref{eqn:hill_diversity} with ${q=1}$ as ${^1D = e^H}$, which is the \emph{award diversity} (RQ2) used herein.

 All of the sampled subreddits have at least one award by kind of post in each dataset, hence the values of ${^1D}$ are within a maximum of Richness $A$ and a minimum of 1. It should thus be noted that, unlike traditional diversity indices, if some proportion of a subreddit’s awards were randomly removed, the Hill diversity decreases by that proportion~\cite{roswell2021conceptual}.

\section{Results}

%TODO: Add introductory paragraph

\subsection{Awarding Levels}

After diversification, the awarding levels increased considerably for all three measures and both post kinds. Based on a paired Wilcoxon signed-rank test at the subreddit level, this increase was significant in all cases (${p \ll .001}$).
Regarding the growth at the dataset level for both post kinds (see Table~\ref{tab:dataset-growth}), awards given was the measure that grew the most, with an average of +175.5\%; then coins spent with an average of +122.5\%; and lastly awarded posts with an average of +121\%.
Hence, the number of coins spent per awards given decreased after diversification: in \dsname{2019H1} it was 381 for submissions and 295 for comments, while in \dsname{2020H1} it was 292 for submissions and 249 for comments, which represents a respective growth of \textminus23.4\% and \textminus15.7\%.
In other words, cheaper awards were the ones that became more popular.
Interestingly, at the dataset level the number of submissions is an order of magnitude lower than the number of comments, but in terms of awarding measures is the opposite: submissions have an awarding level an order of magnitude higher than comments.

\begin{table}
	\centering
	\begin{tabular}[t]{lrrr}
		\toprule
		 & \dsname{2019H1} & \dsname{2020H1} & growth\\
		\midrule
		\addlinespace[0.3em]
		\multicolumn{4}{l}{Submissions}\\
		\hspace{1em}awards given & 6.38 & 18.89 & +196\%\\
		\hspace{1em}coins spent & 2434.50 & 5514.69 & +127\%\\
		\hspace{1em}awarded posts & 3.20 & 6.85 & +114\%\\
		\addlinespace[0.3em]
		\multicolumn{4}{l}{Comments}\\
		\hspace{1em}awards given & 1.24 & 3.15 & +155\%\\
		\hspace{1em}coins spent & 364.90 & 783.17 & +115\%\\
		\hspace{1em}awarded posts & 0.89 & 2.06 & +131\%\\
		\bottomrule
	\end{tabular}
	\caption{\label{tab:dataset-growth}Growth of awarding measures at the dataset level. Measure values are  \textit{per thousand posts}.}
\end{table}

\begin{figure}
   \centering
   \includegraphics[width=1\columnwidth]{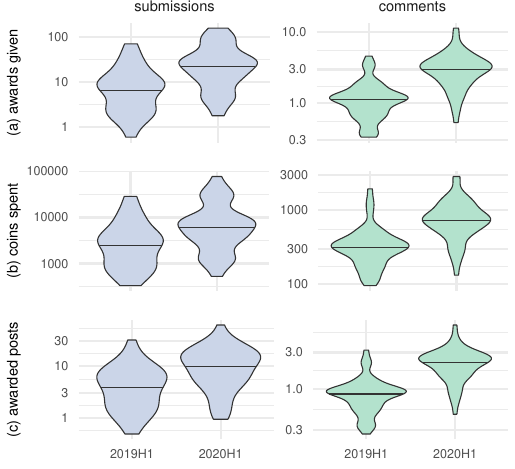}
   \caption{Violin plots of awarding measures at the subreddit level. Median increases after award diversification are notable in all cases. Values are \textit{per thousand posts} on a logarithmic scale.}
   \Description{Figure is divided in three subfigures: (a) awarded posts, (b) coins spent, and (c) awards given; each subfigure is in turn divided in two post kinds: submissions and comments. All subfigures depict a remarkable median increase between 2019H1 and 2020H1 for all measures and post kinds, more so for comments.}
   \label{fig:awarding_measures}
\end{figure}

At the subreddit level, the median values of awarding measures reflect similar considerable increases, albeit the awarding effect sizes after diversification are less spread for comments than submissions (see Table~\ref{tab:subreddit-median}).
Indeed, we can see in Figure~\ref{fig:awarding_measures} that the distribution shapes of the awarding measures for both are noticeably different.
Moreover, for comments we can see that there is a remarkable inversion of skewness after diversification: the distribution on a logarithmic scale goes from positively skewed in \dsname{2019H1} to negatively skewed in \dsname{2020H1}.
In other words, before diversification most subreddits had awarding levels for comments below the subreddit-level mean, while afterward is the contrary. Submission awarding measures also manifest such change in skewness, but in a much less remarkable way.

\begin{figure}
   \centering
   \includegraphics[width=1\columnwidth]{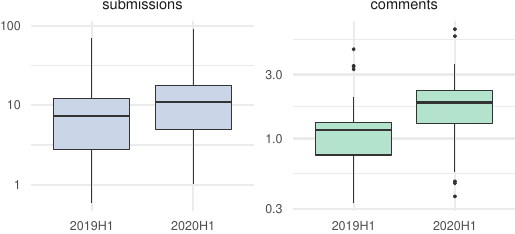}
   \caption{Box plots of gilding-only awards given \textit{per thousand posts} on a logarithmic scale at the subreddit level.}
   \Description{Figure is divided in two post kinds: submissions and comments. Box-plots for both post kinds are similar in their shape and denote an increase after diversification, with it being higher for comments.}
   \label{fig:gilding_measure}
\end{figure}

\begin{figure}
   \centering
   \includegraphics[width=1\columnwidth]{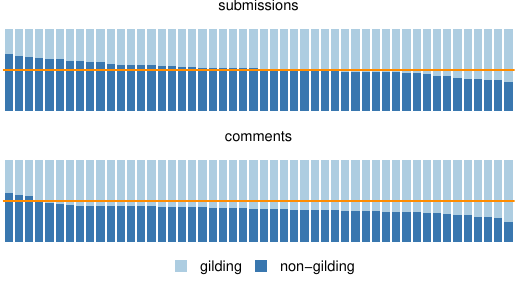}
   \caption{Proportions of awards by gilding group at the subreddit level for \dsname{2020H1}. The orange line is at the 50\% mark.}
   \Description{Figure is divided in two post kinds: submissions and comments. For each, a series of stacked bar-charts shows the proportions of gilding and non-gilded awards for the 50 sampled subreddits. For submissions, the proportions are almost equal across the subreddits; for comments, there are slightly more gilding awards across most subreddits.}
   \label{fig:gilding_status_proportions}
\end{figure}

To discern if the awarding increase was related to the new non-gilding awards and not only to an increase in the old gilding-only awards, the number of awards given by gilding-status was also analyzed.
At the dataset level, the number of gilding-only awards given experiences an increase after diversification, with +31.7\% for submissions and +48.3\% for comments.
At the subreddit level (see Figure~\ref{fig:gilding_measure}), the median awards given \ptp{} in \dsname{2020H} was 11.04 for submissions and 1.86 for comments.
Based on a paired Wilcoxon signed-rank test, the effect is  significant in both cases (${p<.001}$), with a paired median difference of 3.06 with 95\% CI (1.86; 4.32) for submissions, and 0.66 with 95\% CI (0.58; 0.79) for comments.
In \dsname{2020H1}, the 3 gilding-only awards comprised a sizeable portion of the awardings compared to the 450 non-gilding awards. At the dataset level, gilding-only awardings represented 44.5\% for submissions and 58.2\% for comments, with median values at the subreddit level being almost the same.
As illustrated in Figure~\ref{fig:gilding_status_proportions}, of the 50 sampled subreddits, almost half (23) preferred gilding awards for submissions, while this preference extends to the vast majority (46) of subreddits for comments. Naturally, this preference impacts the diversity of awards in \dsname{2020H1}.

\begin{table}
	\centering
	\begin{tabular}[t]{lrrrc}
		\toprule
		 & \multicolumn{2}{c}{Obs. median} & \multicolumn{2}{c}{Paired median diff.}\\
		 \cmidrule(lr){2-3}\cmidrule(lr){4-5}
		 & \dsname{2019H1} & \dsname{2020H1} & diff. & 95\% CI \\
		\midrule
		\addlinespace[0.3em]
		\multicolumn{5}{l}{Submissions}\\
		\hspace{1em}awards given & 7.25 & 22.69 & 14.85 & (11.01; 19.92)\\
		\hspace{1em}coins spent & 2761.91 & 6459.03 & 3590.11 & (2532; 5003)\\
		\hspace{1em}awarded posts & 4.03 & 10.68 & 5.86 & (4.32; 8.26)\\
		\addlinespace[0.3em]
		\multicolumn{5}{l}{Comments}\\
		\hspace{1em}awards given & 1.16 & 3.10 & 1.93 & (1.51; 2.14)\\
		\hspace{1em}coins spent & 322.24 & 759.96 & 445.67 & (323; 485)\\
		\hspace{1em}awarded posts & 0.91 & 2.38 & 1.42 & (1.10; 1.55)\\
		\bottomrule
	\end{tabular}
	\caption{\label{tab:subreddit-median}Estimated median effect sizes \textit{per thousand posts} of diversification on awarding at the subreddit level, based on paired median differences from 5000 bootstrap resamples. Confidence intervals are bias-corrected and accelerated.}
\end{table}

\subsection{Award Diversities}

\begin{figure}
   \centering
   \includegraphics[width=1\columnwidth]{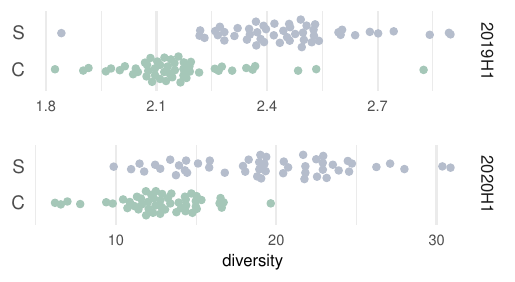}
   \caption{Beeswarm plots of award $^1D$ (diversity) at the subreddit level by submissions (S) and comments (C).}
   \Description{Figure is divided in three subfigures: (a) awarded posts, (b) coins spent, and (c) awards given; each subfigure is in turn divided in two post kinds: submissions and comments. All subfigures depict a remarkable median increase between 2019H1 and 2020H1 for all measures and post kinds, more so for comments.}
   \label{fig:beeswarm_hill_diversity}
\end{figure}

At the dataset level, the diversity ${^1D}$ in \dsname{2019H1} was 2.45 for submissions and 2.17 for comments, while in \dsname{2020H1} it was respectively 28 and 15.2, which represents a growth of +1043.32\% for submissions and +598.95\% for comments. At the subreddit level, in \dsname{2019H1} the median value was 2.45 \mad{0.12} for submissions and 2.13 \mad{0.08} for comments, while in \dsname{2020H1} it was 19.3 \mad{5.36} and 12.8 \mad{2.25}, respectively. The paired median difference was 16.85 with 95\% CI (15.72; 19.25) for submissions, and 10.67 with 95\% CI (10; 11.45) for comments. Hence, and as illustrated in Figure~\ref{fig:beeswarm_hill_diversity}, in both datasets submissions had in average a higher diversity, while also having a higher $^1D$ increase after diversification in both spread and median values with respect to comments. Interestingly, the diversity between both datasets was modestly correlated at the comment level ($\tau = .22$, $p=0.02$) but not at the submission level ($\tau = .04$, $p>0.6$).

\begin{figure}
   \centering
   \includegraphics[width=1\columnwidth]{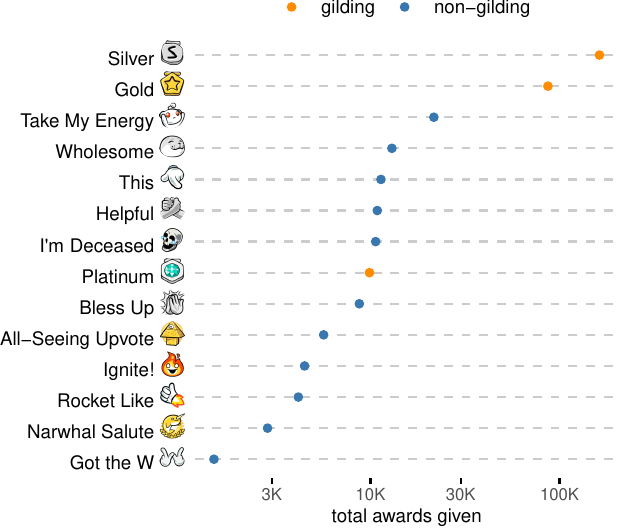}
   \caption{Ranking by total awards given for both post kinds of the 14 awards present in all of the subreddits of \dsname{2020H1}.}
   \Description{Figure is divided in two post kinds: submissions and comments. Box-plots for both post kinds are similar in their shape and denote an increase after diversification, with it being higher for comments.}
   \label{fig:common_awards}
\end{figure}

As mentioned before, in \dsname{2020H1} the old gilding awards had (in average) a higher proportion compared to the new non-gilding awards.
In fact, \award{Silver} and \award{Gold} are respectively the first and second most given awards, both before and after diversification. The three gilding awards are also part of the 14 awards that are present in all of the 50 sampled subreddits, with \award{Silver} and \award{Gold} being at least an order of magnitude higher in terms of total awards given compared to the rest of the awards (see Figure~\ref{fig:common_awards}). It should be noted that, although the proportions of the three gilding awards did not suffer significant changes after diversification, \award{Platinum} (the most expensive gilding award) remained among these 14 common awards, but it fell to the eighth place by total awards given.

\begin{figure*}
   \centering
   \includegraphics[width=1\textwidth]{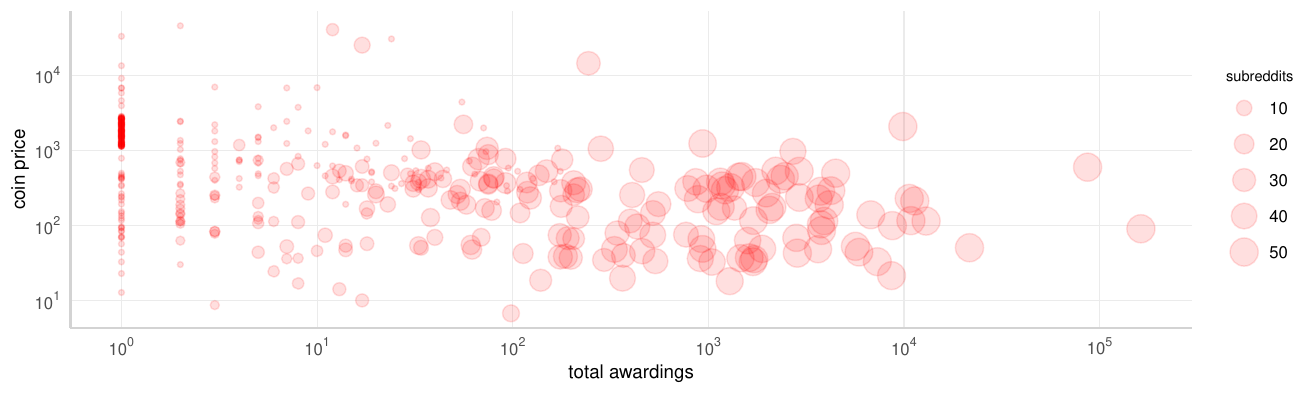}
   \caption{Scatter plot of all of the 453 awards in \dsname{2020H1} by coin price, awardings, and number of subreddits in which the award is present. Jittering was applied at the vertical axis (coin price) to better represent data distribution.}
   \Description{Figure is a bidimensional scatterplot on a logarithmic scale; awardings are on the abscissa, coin price on the ordinate, and the datapoint size is proportional to the number of subreddits in which the award is present. Most awards are only given a few times and are present in a few subreddits, with a coin price close to the median. In general, the higher the number of subreddits the most awarding.}
   \label{fig:all_awards_scatter}
\end{figure*}

In \dsname{2020H1}, there are a total of 453 awards. The median coin price is 500 (mean=1436), with the minimum price being 10 and the maximum 50,000.
Only 69 of these awards (15.23\%) included a coin reward, with a median of 100 coins (mean=434.9), a minimum of 5, and a maximum of 5000.
There was a significant ($p<.01$) moderate positive correlation between subreddit diversity and number of posts for both submissions ($\tau=.22$) and comments ($\tau=.19$).
A considerable number of awards (165) were given only once, with the median being 5 times (mean = 1066). Most awards (257) only appear in one of the fifty subreddits (mean=11.68). The top three awards by their given times were \award{Silver} (162,652 times), \award{Gold} (86,830 times), and \award{Take My Energy} (21,625 times). This data distribution is illustrated in Figure~\ref{fig:all_awards_scatter}.

\section{Discussion}

\subsection{The more the better}

The analysis results indicate that the diversification of Reddit Awards attained its goal of increasing platform-wise awarding. In average, all of the post awarding measures had a growth of +140\% in one year.
As a comparison, in the pre-Coins gilding study mentioned in \S\ref{sec:background} with the top 100 subreddit by number for comments for the month of May of 2015, the number of gilded comments at the dataset level was of 0.411 \ptp{}~\cite{mendelsohn2017giving}.
Thus, in circa four years ---and after an extension to three awards--- there was a growth of only +116\% for gilded comments, while the equivalent one-year increase after diversification for awarded comments was of +131\% (see Table~\ref{tab:dataset-growth}).

Concerning the diversity of awards, the increase was important but not that much noticeable as one could have expected, albeit submissions had more noticeable increase in diversity compared to comments. Most awards in the dataset were given only a very few times, although some other new awards found great success, as illustrated in Figure~\ref{fig:common_awards}.
Indeed the growth in awarding was not all thanks to the new awards. The three gilding awards also grew slightly, and both \award{Silver} and \award{Gold} continued to be the most popular awards.
There might be a few reasons for the popularity of these awards.
There is the price and perks; \award{Silver} costs 100 coins and has no perks; \award{Gold} costs 500 coins and includes a reward of 100 coins and a week of ad-free browsing; and, for completeness, the coin price of \award{Platinum} is 1800, with ad-free browsing and 700 coins as reward, both valid only for a month.

Gold was the first and original award, and at the moment of writing is the default award upon awarding (see Figure~\ref{fig:award_modal}), thus there is a sense of familiarity; the same, but to a lesser extent as with \award{Silver}. As a matter of fact, \award{Silver} had its origin as an inside joke among users. It was an image of a poorly drawn and engraved silver medal, which was linked on comments as a humorous statement that the content was appreciated but not sufficiently for \award{Gold} or as a token of appreciation for which the ``giver'' had no sufficient funds. Reddit flew with the idea and made the \award{Silver} award for the Reddit Coins introduction, retaining the original crude drawing and the lack of benefits, which has now become by far the most given award platform-wise. The case of \award{Silver} highlights the fact that users might be driven to use a particular award by its meaning (explicit or implicit), rather than by its price or perks.

\subsection{Limitations}

This study has several limitations, all of which might inspire future work on the subject. The sampling method might have given biased results at the subreddit level; perhaps smaller subreddits are more or less generous in their awarding, although at the dataset level the sampled subreddits represent an important and diverse share of the platform's content.
Similarly, the less numerous yet important category of NSFW subreddits might have resulted in different awarding dynamics due to its mature content. For instance, \subr{gonewild}, the only NSFW community present in the initial top 50 rank (see \S\ref{sec:data}), has some peculiarities.
According to a pre-Coins study on the subreddit~\cite{lloyd2016porno}, circa 25\% of a sample of 90 females users (the vast majority of submissions are by women) with 3,454 individual photos was gilded at least once, and had 23 comments per photo (the vast majority by men). These female users were chosen evenly by both account age and activity, although no more details are given regarding content gilding. However, I think that this NSFW subreddit is an outlier, taking into consideration its amateur nature, sexually provoking dynamic among users, and immense popularity (it is the only such subreddit ranked in the top lists).

Another sampling limitation concerns the timeframe of the data collected; given that only a few months had passed after diversification in \dsname{2020H1}, perhaps the proportion between gilding and non-gilding content could have been different for a later time period, albeit unrelated exogenous phenomena might had confounded the results.
Endogenous phenomena might have also affected the study results. In particular, the graphical user interface might have a significant impact on which awards are chosen, if any.
As can be seen in Figure~\ref{fig:award_modal}, during award selection there are several tabs which group the awards, and only a small subset of the awards is immediately visible upon opening the selection dialog.
It is most likely that the item disposition in this section is carefully monitored, since a few years Reddit revamped its approach to graphical design and user interaction.\footnote{\url{https://www.wired.com/story/reddit-redesign/}}

\subsection{Implications for social media platforms}

Overall, results offer strong supporting evidence for the benefits of diversifying paid appreciation tokens, which might represent an advisable monetization strategy of user-generated content for other social media platforms that already use paid appreciation tokens or are interested in them.
In particular, the case of Reddit Awards is of relevance to platforms centered around user-generated content for and by peers, such as discussion forums and question-and-answers websites.
Moreover, peer awards could complement or substitute existing achievement badges ---which are automatically given by the system and are non-paid--- such as those present in the platform Stack Overflow as way to gamify and engage the question and answer nature of its user-generated content~\cite{marder2015stack}.
Careful attention should be paid to user involvement in the design and development of peer award mechanisms, however.

Indeed, significant changes in the design of a peer user-generated content platforms sometimes might have catastrophic consequences, such as with the redesign of the question-and-answers AnswerBag in 2009~\cite{gazan2011redesign} and the infamous fourth version of the news aggregator Digg in 2010~\cite{germonprez2013member}, with the latter provoking a mass migration towards Reddit and even becoming a meme.\footnote{\url{https://knowyourmeme.com/memes/events/digg-v4}}
This event was in fact one of the main drivers for the aforementioned revamped design approach of Reddit.
Important new features such as Reddit Awards have now a period of pilot testing and community feedback before their platform-wise deployment so as to lead to a successful adoption by users.
Furthermore, Reddit Awards have been extensively promoted and are considered to be key for the future of the platform, as stated by Reddit administrator \redditor{venkman01} in an official company statement:\footnote{\url{https://redd.it/hmdwxs}}

\begin{quote}
Awarding is an important part of our direct-to-consumer revenue; it complements advertising revenue and gives us a strong footing to pursue our mission into the future. By giving awards, users not only recognize others but also help Reddit in its mission to bring more community and belonging to the world.	
\end{quote}

It is likely that this future also includes the offering of Reddit Awards (or other paid appreciation tokens) as digital property via the use of non-fungible tokens (NFTs) and blockchain technology. An NFT is a unique identifier recorded in a distributed ledger (e.g., a blockchain) that can be used to certify the  ownership of digital objects, which have recently gathered much hype due to high-profile sales~\cite{trujillo2022surge}. In fact, Reddit has recently started to experiment with the selling of NFTs, in the form of unique profile avatars.\footnote{\url{https://nft.reddit.com}} I thus foresee that, despite the legal issues with virtual currency and goods mentioned in \S\ref{sec:background}, in the near future social media platforms will at least explore expanding their revenue streams by going to the extreme of diversification and offering paid appreciation non-fungible tokens, selling the idea of ``a unique object for a unique content recognition''.

\section{Conclusion}

Paid appreciation tokens have become one of the many strategies with which social media platforms are retaining and monetizing user-generated content. This work focused on the  diversification effects of Reddit Awards, a peer awarding mechanism in which any user can be content creator and content consumer at the same time, without the expectation of financial gain. To this end, a set of measures to characterize and operationalize awarding and diversification were defined and analyzed, which could be further adapted to investigate similar paid appreciation tokens in other platforms. Results indicate that providing more award options brings a benefit to users ---who obtain entertainment seeing others post awards, as well as satisfaction in receiving or giving awards--- and the platform ---which increases its revenue. Still, already familiar awards remained highly popular, thus attention should be paid to these when introducing changes. Being one of the few studies on the topic, the current work has several limitations. Further research is thus needed to better understand the user design implications of peer awarding mechanisms in social platforms.

\bibliographystyle{ACM-Reference-Format}
\bibliography{bibliography.bib}

\end{document}